# High-average-power femtosecond laser at 258 nm

MICHAEL MÜLLER,[1,*] ARNO KLENKE,[1,2] THOMAS GOTTSCHALL,[1] ROBERT KLAS,[1] CAROLIN ROTHHARDT,[1,3] STEFAN DEMMLER,[1,2] JAN ROTHHARDT,[1,2] JENS LIMPERT[1,2,3], ANDREAS TÜNNERMANN[1,2,3]

[1]Friedrich-Schiller-University Jena, Abbe Center of Photonics, Institute of Applied Physics, Albert-Einstein-Straße 15, 07745 Jena, Germany
[2]Helmholtz-Institute Jena, Fröbelstieg 3, 07743 Jena, Germany
[3]Fraunhofer Institute for Applied Optics and Precision Engineering, Albert-Einstein-Straße 7, 07745 Jena, Germany
*Corresponding author: michael.mm.mueller@uni-jena.de

**We present an ultrafast fiber laser system delivering 4.6 W average power at 258 nm based on two-stage fourth-harmonic generation in beta barium borate (BBO). The beam quality is close to diffraction limited with an M$^2$-value of 1.3 x 1.6. The pulse duration is 150 fs, which potentially is compressible down to 40 fs. A plain BBO and a Sapphire-BBO compound are compared with respect to the achievable beam quality in the conversion process. This laser is applicable in scientific and industrial fields. Further scaling to higher average power is discussed.**

Today, there is an increasing interest in ultrafast ultraviolet (UV) lasers with high beam quality for scientific and industrial applications. For example, precision machining benefits from the short wavelength combined with the short pulse duration as it enables smaller feature sizes and small heat affected zones [1]. Moreover, the increased linear absorption simplifies the machining of large bandgap materials [2].

Scientific applications include UV induced breakdown spectroscopy [3], ultrafast spectroscopy [4] and generation of extreme ultraviolet (XUV) light via high-harmonic generation (HHG). For HHG, the short driving wavelength leads to a significant increase of the conversion efficiency compared to state-of-the-art infrared (IR) driving wavelength [5–8]. The ultrafast UV lasers required for this process are usually made by frequency doubling or tripling of femtosecond IR lasers in nonlinear crystals, e.g. beta barium borate (BBO). The crystals have to be very thin to allow a suitable phase matching bandwidth. Following this approach, the second harmonic (SH, 515 nm) of an Ytterbium-based ultrafast fiber laser has been used to generate coherent XUV light via HHG at an average power of 800 µW and at a photon energy of 21.7 eV with an Green-to-XUV conversion efficiency of 7.5·10$^{-5}$ [9]. A further increase in conversion efficiency and XUV power is desired by 'photon hungry' applications such as nanoscale imaging [10,11], studies on molecular dynamics [12] or spectroscopy of the energy levels of highly-charged ions [13].

This increase in photon flux will be possible by using even shorter driving wavelengths and even more input power. However, generating high-power ultrashort pulses with high beam quality in the UV is challenging, as absorption in the nonlinear crystals leads to beam quality degradation and even crystal fracture. The state of the art in frequency converted IR lasers are 2.6 W of average power with 35 fs pulse duration, but without a measurement of the beam quality [6] and 2.7 W average power with an M$^2$-value of 2.5 at 4.2 ps pulse duration [14]. The frequency conversion can also be accomplished in gas-filled waveguides with fairly low conversion efficiency [15]. Furthermore, there are KrF excimer [16] and Ce:LiCAF [17] lasers delivering femtosecond UV pulses with up to 50 W average power, but at very low repetition rates and unspecified beam quality.

Following the approach of frequency conversion, a 100 W average power third-harmonic femtosecond laser at 343 nm with diffraction limited beam quality has been demonstrated recently [18]. In this system, a sandwich structure consisting of a BBO bonded between highly thermoconductive sapphire plates is used to mitigate the detrimental thermal issue in the BBO.

In this paper, we extend this work towards shorter wavelength and present a nearly diffraction limited high-power ultrafast laser at 258 nm based on the fourth harmonic (FH) of an Yb-doped ultrafast fiber laser system. We compare the power and the beam quality of the UV beam generated in a plain BBO and in a Sapphire-BBO-Sapphire sandwich structure. The challenge is to overcome one order of magnitude larger linear and almost two orders of magnitude larger nonlinear absorption in both BBO and sapphire at 258 nm compared to the existing 343 nm laser system [19,20].

The experimental setup for the two-stage fourth-harmonic generation is depicted in Fig. 1. First, the output of an ultrafast fiber chirped-pulse amplification system [21] is directed through a half-wave plate and a thin-film polarizer to control the input power to the experiment while the laser is operated at maximum output power. The beam is collimated to 1.2 mm $1/e^2$-diameter by means of a Galilean telescope (lenses $f_1$ and $f_2$). The beam is sent through a 0.5 mm thick BBO crystal cut at 23.4° for type I phase matched second-harmonic generation. The crystal is anti-reflective coated for both the fundamental and the second harmonic and is mounted in a tip, tilt and rotation mount to optimize phase matching conditions. The two wavelengths are separated by means of two 45° dichroic mirrors and the green light is used to generate the fourth harmonic in another type-I-phase-matched BBO cut at 50.0°. Two types of BBO crystals were used to compare their performance in the experiment. First, an uncoated 0.1 mm thick BBO, was implemented. Then a sapphire-BBO-sapphire sandwich structure consisting of two 1 mm thick sapphire plates and a 0.1 mm thick BBO was used. The sandwich structure was fabricated via direct bonding and the outer surfaces are anti-reflective coated for both the SH and FH. This compound-crystal was held in a water-cooled mount with only tilt-adjustment. Thus, a half-wave plate to rotate the polarization of the green driving pulse is used additionally to achieve phase matching. After the fourth harmonic is generated, it is separated from the remaining second harmonic using two dichroic mirrors at 45° angle of incidence. A sequence of two sapphire wedges at Brewster angle is used to take an s-polarized sample of the UV beam to measure the pulse duration, the beam quality, and the spectrum. A second-order cross-correlation based on difference frequency generation of the fourth harmonic with the fundamental wave is used to measure the pulse duration. Thus, a beam sample of the infrared driving field is taken prior the frequency conversion. An optical delay stage is used to match the propagation time of the infrared and the UV pulse to the cross-correlator.

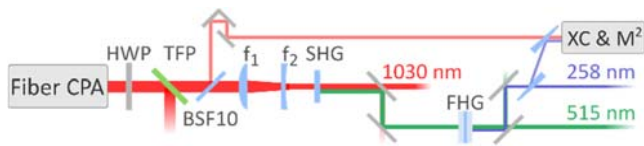

Fig. 1. Experimental setup of the two-stage fourth harmonic generation. HWP: half-wave plate, TFP: thin-film polarizer, $f_1$ and $f_2$: telescope, SHG: second-harmonic crystal, FHG: fourth-harmonic crystal, XC: cross-correlator, $M^2$: beam quality measurement.

At the beginning of the experiment, the infrared beam is characterized. The fiber laser is operated at 80 W average power and at a repetition rate of 796 kHz. The optical spectrum, as shown in Fig. 2, is centered at 1032 nm with a full width at half maximum (FWHM) of 10.2 nm. The high-frequency ripples are due to a spatial light modulator in the fiber-laser system and the slow modulations are due to a spectral amplitude shaper. The corresponding second-order auto-correlation is shown in Fig. 3 and features a FWHM of 301 fs, which corresponds to a pulse duration of 213 fs assuming a Gaussian pulse. Furthermore, the $M^2$-value is measured ($4\sigma$-method) and is found to be less than 1.1 on both axes of the beam, which is consistent with previously reported values [21].

Now, the infrared beam is directed into the frequency conversion stages. The crystal orientations of the SHG and FHG stages are optimized to generate the highest power in the UV, which is 7.5 W. The $M^2$-value of the second harmonic is measured to be less than 1.1 on both axes at 40 W average power. Then, the $M^2$-measurement is done for the fourth harmonic as well. The results for the plain BBO is shown in Fig. 4 for increasing UV power. The initial $M^2$-value is 1.4 on both axes and remains at this level up to approximately 4 W of UV. Then, heating of the crystal due to linear and nonlinear UV absorption leads to a significant thermal lens, distorting the beam, such that at 7.5 W UV, the $M^2$-value increases up to 3.4.

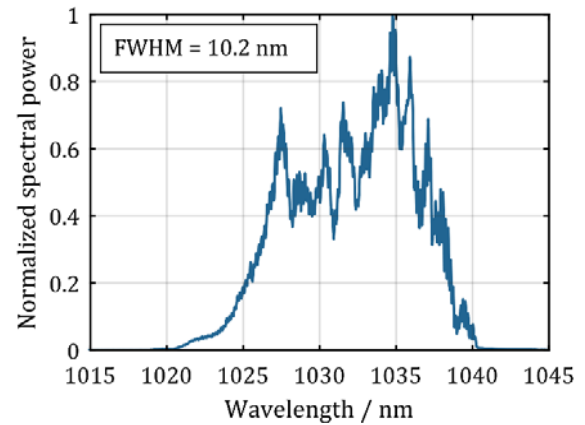

Fig. 2. Optical spectrum of the infrared input pulse.

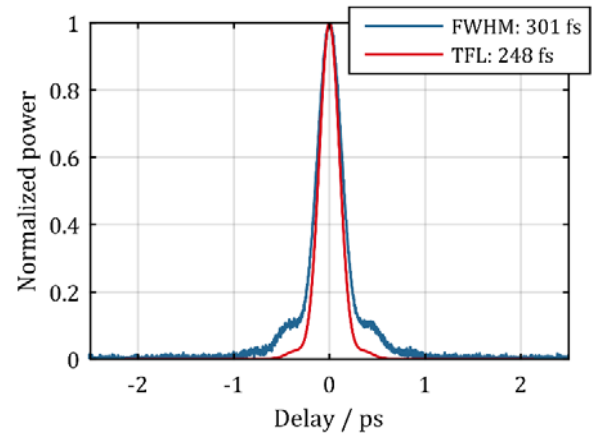

Fig. 3. Measured, background-free autocorrelation trace and calculated, transform-limited autocorrelation trace of the fiber CPA.

The experiment is set to 4.6 W of UV and both the UV output and the green driver (prior FHG) are analyzed in more detail. The optical spectra were measured with 0.1 nm resolution from stray light of the power meters or beam dumps. The second harmonic spectrum Fig. 5(a) has a FWHM of 2.6 nm, supporting a 130 fs pulse. This is a realistic result, as in SHG generation the pulse duration reduces by a factor of $\sqrt{2}$ for a Gaussian input pulse, leading to such an optical bandwidth. The FHG spectrum has an increased bandwidth with a

FWHM of 3.1 nm and is depicted in Fig. 5(b). A possible reasons for the bandwidth increase is nonlinear phase accumulation in the BBO crystals leading to a small pulse chirp that can result in spectral broadening during the conversion.

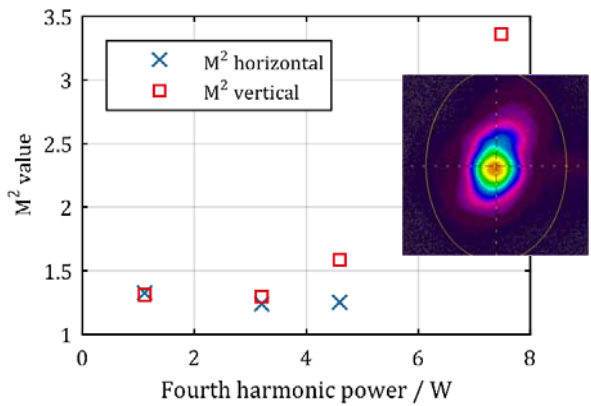

Fig. 4. Measured $M^2$-value of the UV beam versus the generated UV power from the 0.1mm BBO. Inset: Beam focus at 4.6W UV power.

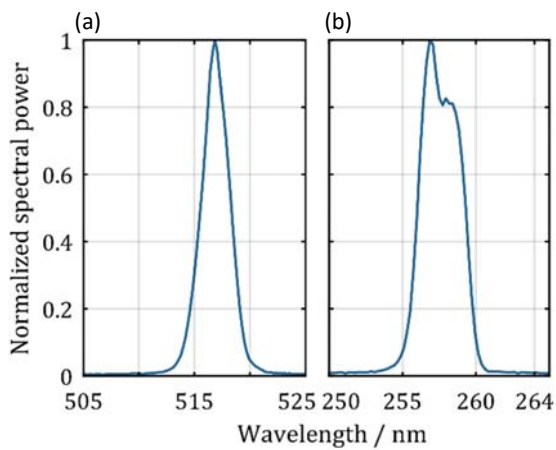

Fig. 5. Optical spectra of (a) the second harmonic and (b) the fourth harmonic (b).

Next, the cross-correlation of the UV-pulse and the IR-driver is measured, which is shown in Fig. 6. The FWHM duration of the UV pulse is 150 fs as calculated from the 264 fs FWHM of the cross-correlation and from the initially determined 213 fs FWHM pulse duration of the IR driver. In future, compression to the transform limit of the UV spectrum with 40 fs pulse duration might be feasible, e.g. by using a prism pair or chirped mirrors.

After these investigations, the plain BBO is replaced by the Sapphire-BBO-Sapphire compound in the FHG stage. The compound should offer an improved heat dissipation and allow to further increase the UV average power without loss of beam quality. The experimentally determined $M^2$-values, as depicted in Fig. 7, are disproving this assumption. The $M^2$-value at low average power already is in the range of 1.6 to 1.8 on both axes. At 3 W UV power, the $M^2$-value increases to almost 2. At even higher UV power, the beam quality degrades rapidly, such that the $M^2$-measurement yields no satisfactory fit to the beam caustic any more. Hence, in terms of beam quality, the sapphire-BBO-sapphire sandwich structure does not perform better than a plain piece of BBO. This contradicts the experience made in THG [18] and is investigated in the following

First, the surface temperature of the sandwich structure is monitored using a thermographic camera, which is corrected for the emissivity of the sapphire plate ($\varepsilon_{Sapphire}$ = 0.95 [22]). The measurement reveals a temperature gradient of 3°C from the beam axis to substrate's edge at 4 W UV average power. Then, the sapphire-BBO compound is replaced with the plain BBO and again 4 W of UV are generated. A 10x10x1 mm sapphire plate, which is identical to the compound's heat spreaders, is placed prior and after the BBO crystal. The surface temperatures of both BBO and sapphire are monitored and corrected for their emissivity ($\varepsilon_{BBO}$ = 0.78, [23]). No temperature increase in the sapphire plate is observed when it is placed in the green beam before the FHG-BBO – hence absorption of green light expectedly is not an issue. However, when the plate is placed in the UV beam, its temperature rises to 43°C on the beam axis and decreases to 39°C on the substrate edge – the same gradient as observed on the sandwich structure. In this setting, the $M^2$-value of the UV beam increases by 0.1-0.2 on both axes due to the thermal lens in the sapphire plate. The origin of this thermal load is both linear and two-photon absorption (TPA). The TPA absorption is $9.4 \cdot 10^{-11}$ cm/W [19], which yields a TPA absorption coefficient of 0.17 cm$^{-1}$ at an estimated peak intensity of $2 \cdot 10^9$ W/cm$^2$ for 1mm beam diameter. This contribution is about equal to the linear absorption coefficient of 0.19 cm$^{-1}$ at 258 nm [24]. Thus, the heat input in the sapphire heat spreaders is significantly increased compared to the cases of SHG and THG and explains the observed temperature gradient.

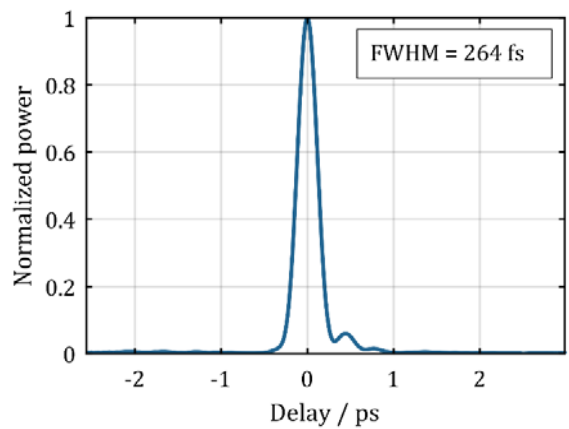

Fig. 6. Cross-correlation of the fundamental pulse with the fourth harmonic indicating a 150 fs pulse duration of the UV pulse.

At the same time, the surface temperature of the plain BBO is 165°C on axis and 45°C at the substrate edge. The large temperature gradient observed is due to the almost one order of magnitude stronger linear [25] and nonlinear absorption (UV TPA: $68 \cdot 10^{-11}$ cm/W [19] and UV-green-absorption [26]) to plus the approximately one order of magnitude lower thermal conductivity compared to sapphire. Thus, the main contribution to the beam quality degradation at high UV power still is the BBO itself. The sapphire plates will still improve on the heat dissipation of the BBO and reduce its thermal lensing, but this effect is not overcoming the additional thermal lens in the sapphire.

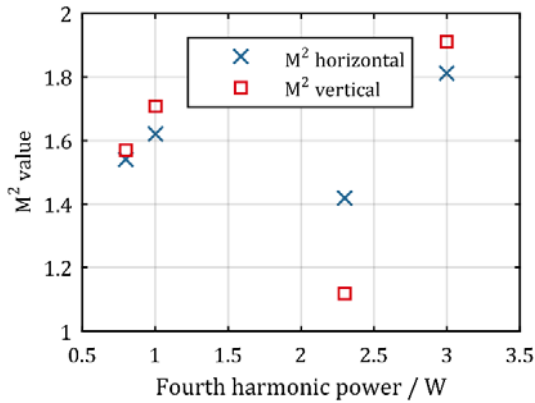

Fig. 7. Measured $M^2$-value of the UV beam versus the generated UV power from the Sapphire-BBO-Sapphire stack.

The significant increase of both linear and nonlinear absorption impedes power scaling using the sandwich structure compared to the previous SHG and THG experiments, where the compound outperformed the plain crystal [18,23]. However, the concept is still viable in FHG for ns-pulsed and continuous-wave lasers, as the nonlinear absorption is much weaker in that case. Then, crystal stacks with several layers have to be manufactured to achieve sufficient conversion length, which is not possible in the femtosecond regime due to temporal walk-off and dispersion.

In summary, we presented a high-power ultrafast UV laser at 258 nm based on two-stage fourth-harmonic generation of a 1 µm ultrafast fiber laser. The achieved UV output parameters are 4.6 W average power at 258 nm with 150 fs pulse duration. The beam quality is close-to-diffraction-limited with M2 = 1.3 x 1.6. These performance figures overcome the state-of-the-art average power by a factor 2 [6]. We compared a plain BBO to a Sapphire-BBO-Sapphire sandwich structure for the UV-generation. The advantage of the sapphire heat spreaders as in third-harmonic generation is not given for fourth-harmonic generation, due to significantly increased linear and nonlinear absorption of the generated UV in the sapphire heat-spreader plates.

Future work will focus on increasing the UV average power by using CLBO or periodically poled LBO crystals, which have a higher UV absorption edge. Moreover, spatially inhomogeneous cooling will be applied [27] and the compression of the UV pulses will be pursued., enabling 10 W-class femtosecond UV lasers with diffraction limited beam quality for a broad range of industrial and scientific applications.

**Acknowledgment**. We acknowledge funding by the German Ministry of Education and Research (Bundesministerium für Bildung und Forschung, BMBF), grant number 05P15SJFFA. M.M. acknowledges financial support by the Carl-Zeiss-Stiftung.

## References

1. J. H. Klein-Wiele, J. Bekesi, and P. Simon, Appl. Phys. A **79**, 775–778 (2004).
2. A. Saliminia, A. Proulx, and R. Vallée, Opt. Commun. **333**, 133–138 (2014).
3. S. P. Banerjee, Z. Chen, and R. Fedosejevs, Opt. Lasers Eng. **68**, 1–6 (2015).
4. T. Kobayashi, A. Yabushita, and Y. Kida, Photonics **3**, 64 (2016).
5. A. D. Shiner, C. Trallero-Herrero, N. Kajumba, H. C. Bandulet, D. Comtois, F. Légaré, M. Giguère, J. C. Kieffer, P. B. Corkum, and D. M. Villeneuve, Phys. Rev. Lett. **103**, 1–4 (2009).
6. D. Popmintchev, C. Hernández-García, F. Dollar, C. Mancuso, J. A. Pérez-Hernández, M.-C. Chen, A. Hankla, X. Gao, B. Shim, A. L. Gaeta, M. Tarazkar, D. A. Romanov, R. J. Levis, J. A. Gaffney, M. Foord, S. B. Libby, A. Jaron-Becker, A. Becker, L. Plaja, M. M. Murnane, H. C. Kapteyn, and T. Popmintchev, Science **350**, 1225–31 (2015).
7. H. Wang, Y. Xu, S. Ulonska, J. S. Robinson, P. Ranitovic, and R. A. Kaindl, Nat Commun **6**, (2015).
8. D. Popmintchev, M.-C. Chen, C. H. García, J. A. P. Hernández, J. P. Siqueira, S. Brown, F. Dollar, B. C. Walker, P. Grychtol, L. Plaja, M. M. Murnane, H. Kapteyn, and T. Popmintchev, in *CLEO: 2013* (Optical Society of America, 2013), p. QW1A.5.
9. R. Klas, S. Demmler, M. Tschernajew, S. Hädrich, Y. Shamir, A. Tünnermann, J. Rothhardt, and J. Limpert, Optica **3**, 1167 (2016).
10. G. K. Tadesse, R. Klas, S. Demmler, S. Hädrich, I. Wahyutama, M. Steinert, C. Spielmann, M. Zürch, A. Tünnermann, J. Limpert, J. Rothhardt, and Others, Opt. Lett. **41**, 2–6 (2016).
11. A. Ravasio, D. Gauthier, F. R. N. C. Maia, M. Billon, J. P. Caumes, D. Garzella, M. Géléoc, O. Gobert, J. F. Hergott, A. M. Pena, H. Perez, B. Carré, E. Bourhis, J. Gierak, A. Madouri, D. Mailly, B. Schiedt, M. Fajardo, J. Gautier, P. Zeitoun, P. H. Bucksbaum, J. Hajdu, and H. Merdji, Phys. Rev. Lett. **103**, 1–5 (2009).
12. D. Davydova, A. De La Cadena, S. Demmler, J. Rothhardt, J. Limpert, T. Pascher, D. Akimov, and B. Dietzek, Chem. Phys. **464**, 69–77 (2016).
13. J. Rothhardt, S. Hädrich, S. Demmler, M. Krebs, D. F. A. Winters, T. Kühl, T. Stöhlker, J. Limpert, and A. Tünnermann, Phys. Scr. **T166**, 14030 (2015).
14. C. Chang, P. Krogen, H. Liang, G. J. Stein, J. Moses, C. Lai, J. P. Siqueira, L. E. Zapata, F. X. Kärtner, and K. Hong, Opt. Lett. **40**, 665 (2015).
15. L. Misoguti, S. Backus, C. G. Durfee, R. Bartels, M. M. Murnane, and H. C. Kapteyn, Phys. Rev. Lett. **87**, 013601/1-013601/4 (2001).
16. Y. Nabekawa, D. Yoshitomi, T. Sekikawa, and S. Watanabe, IEEE J. Sel. Top. Quantum Electron. **7**, 551–558 (2001).
17. Z. Liu, T. Kozeki, Y. Suzuki, N. Sarukura, K. Shimamura, T. Fukuda, M. Hirano, and H. Hosono, Opt. Lett. **26**, 301–3 (2001).
18. J. Rothhardt, C. Rothhardt, M. Müller, A. Klenke, M. Kienel, S. Demmler, T. Elsman, M. Rothhardt, J. Limpert, and A. Tünnermann, Opt. Lett. **41**, 1885–1888 (2016).
19. A. Dragonmir, J. G. McInerney, and D. N. Nikogosyan, Appl. Opt. **41**, 4365–4376 (2002).
20. R. DeSalvo, A. A. Said, D. J. Hagan, E. W. Van Stryland, and M. Sheik-Bahae, IEEE J. Quantum Electron. **32**, 1324–1333 (1996).
21. M. Müller, M. Kienel, A. Klenke, T. Gottschall, E. Shestaev, M. Plötner, J. Limpert, and A. Tünnermann, Opt. Lett. **41**, 3439 (2016).
22. S. G. Kaplan and L. M. Hanssen, Proc. SPIE **3425**, 120–125 (1998).
23. C. Rothhardt, J. Rothhardt, A. Klenke, T. Peschel, R. Eberhardt, J. Limpert, and A. Tünnermann, Opt. Mater. Express **4**, 1092 (2014).
24. N. A. Kulagin and L. A. Litvinov, Cryst. Res. Technol. **20**, 1667–1672 (1985).
25. J. Liebertz and S. Stähr, Zeitschrift für Krist. Mater. **165**, 91–94 (1983).
26. S. Wu, G. A. Blake, S. Sun, and H. Yu, Proc. SPIE **3928**, 221–227 (2000).
27. Y. K. Yap, K. Deki, N. Kitatochi, Y. Mori, and T. Sasaki, Opt. Lett. **23**, 1016–1018 (1998).